\documentclass[12pt]{article}
\usepackage{amsmath,amssymb,amsthm,color}
\usepackage{graphicx}
\usepackage{cite}
\usepackage{epsfig}
\usepackage{lineno}
\renewcommand{\baselinestretch}{1.66}

\begin{document}
\title{Feature-rich plasmon excitations in sliding bilayer graphene \\}
\author{
\small Chiun-Yan Lin$^{*a}$, Chih-Wei Chiu$^{*b}$, Ming-Fa Lin$^{a,c,d}$ \\
\small $^{a}$Department of Physics, National Cheng Kung University, Taiwan\\
\small $^{b}$Department of Physics, National Kaohsiung Normal University, Kaohsiung, Taiwan\\
\small $^{c}$Hierarchical Green-Energy Materials Research Center, National Cheng Kung University, Tainan, Taiwan\\
\small $^{d}$Quantum topology center, National Cheng Kung University, Tainan, Taiwan\\
 }
\renewcommand{\baselinestretch}{1.66}
\maketitle

\renewcommand{\baselinestretch}{1.66}

\begin{abstract}

This article investigates the intriguing electronic excitations that occur in sliding bilayer graphene. To explore this phenomenon, the study uses a tight-binding model and a modified layer-based random-phase approximation to delve into the complex electron-electron interactions and many-particle excitations involved. By examining the essential electronic properties, such as the density of states, and band structures of bilayer graphene, the study gains a deeper understanding of the elementary excitation phenomena that occur when two graphene layers are shifted relative to each other, transitioning from AA to AB to AA$^\prime$ stacking configurations. The process yields two types of plasmon modes, accompanied by complex single-particle excitations that arise from the distortion and hybridization of the two Dirac cones. The lower-symmetry system is particularly intriguing, exhibiting more complicated Coulomb excitation phenomena. The theoretical predictions can be verified through high-resolution experimental examinations, providing a solid foundation for further research in this exciting field.

\end{abstract}
\par\noindent ~~~~$^*$Corresponding author- E-mail: cylin@mail.ncku.edu.tw (C. Y. Lin)
\par\noindent ~~~~$^*$Corresponding author- E-mail: giorgio@mail.nknu.edu.tw (C. W. Chiu)

\pagebreak
\renewcommand{\baselinestretch}{2}
\newpage

\vskip 0.6 truecm

\section{Introduction}

Bilayer graphene has received a great deal of attention due to its extraordinary electronic and optical properties, which are greatly influenced by the misalignment between layers. In addition to the two standard configurations of bilayer graphene (hexagonal (AA) and Bernal (AB) stacking), researchers have investigated the effects of manipulating the geometric structure, such as twist angles, sliding layers, and domain walls\cite{NatComm12;2391,AdvMater33;2004974,ACSNANO7;1718,PRL108;205503,NatNanotech13;204}, which can induce interband excitations due to the environmental potential of the hexagonal lattice\cite{Nature605;63,NanoLett16;6844,SciRep4;7509}.
In particular, even a slight misalignment between adjacent layers can dramatically alter electronic behavior\cite{SciRep4;7509,APL101;083101,PRB84;155410,SciRep5;17490,
PNAS108;12233,PRB99;205134,JPCC119;10623,PRL100;036804,SciRep9;859}. Previous studies have demonstrated that domain walls alter the environment of normally stacked graphene, resulting in 1D phenomena in electronic and optical properties that rely on the modulated period of the stacking configuration, such as AB/domain wall/BA/domain wall\cite{SciRep9;859,PRL100;036804}.

Misalignment of twisted bilayer graphene induces exotic electron behavior in bilayer graphene, including undamped interband plasmons that are tunable but have a nearly constant energy dispersion in the undoped material, especially for relatively small twist angles ($\theta<2^{\circ}$) that create weakly dispersed flat energy bands\cite{Nature605;63,NanoLett16;6844}. Meanwhile, Mori\'{e} superconductivity, linked to magic angles of approximately 1.1$^{\circ}$ in twisted bilayer graphene, is a phenomenon that has been extensively studied in recent years\cite{Nature556;7699}. It has been found that at a critical temperature of up to 1.7 kelvin, this type of superconductivity can emerge in twisted bilayer graphene with the appropriate angle, leading to interesting and potentially useful properties\cite{AdvSci9;2103170,SCIENCE377;1538}. In this work, sliding bilayer graphene has been shown to exhibit feature-rich plasmon excitations through a relative shift of two graphene layers.
The sliding configuration results in the hybridization of low-lying subbands, leading to two pairs of 2D energy bands and a novel Coulomb excitation spectrum. The unique features observed are attributed to a structural effect on the local atomic environment due to the coordination of the graphene sublattices. These plasmon excitations are particularly important for understanding the interlayer coupling on electron-electron interactions in sliding bilayer graphene and can provide insights into electronic properties such as band anticrossing, band distortion, and collective excitations.

Sliding bilayer graphene, characterized by a continuous variation of the relative displacement between two graphene sheets, has been the subject of numerous physical property investigations. Intermediate configurations between AA and AB stackings exhibit anomalous optical phonon splitting\cite{ACSNano7;7151} and unusual electronic transmission\cite{PRB89;085426,PRB88;115409,CARBON99;432}. Recent first-principle calculations have identified the dependence of phonon frequency on polarized Raman scattering intensity as a signature of tiny misalignments in bilayer graphene\cite{ACSNano7;7151}. While most pristine bilayer graphene displays semimetallic behavior, sliding bilayer graphene exhibits unusual optical excitations at low energies due to the unusual van Hove singularities in the density of states.

In addition, under a uniform perpendicular magnetic field, sliding bilayer graphene induces quantized Landau levels, leading to unusual anti-crossing behavior in the energy spectra and magneto-optical selection rules that differ significantly from those of monolayer and AA- and AB-stacked bilayer systems\cite{IOPBook;978,SciRep4;7509}. The local structure of sliding bilayer graphene, influenced by the geometric symmetry induced by layer displacements in the bilayer system, markedly affects the wave function distribution and leads to unusual energy bands that differ from those in the AA and AB regions. In this study, we aim to comprehensively investigate the Coulomb excitation phenomena of sliding bilayer graphene, with a particular focus on the elementary excitations related to various stacking configurations and the doping effects. It has the potential for applications of sliding bilayer graphene in optoelectronics and plasmodics.

This study employs the tight-binding model and modified random phase approximation (RPA) to investigate the single-particle and many-particle properties of sliding bilayer graphene. Using an empirical formula for the coordinate-dependent hopping integrals\cite{SciRep4;7509}, the study thoroughly explores the essential band structures and DOS as the bilayer graphene transitions from AA to AB to AA$^\prime$ stacking configurations by sliding one layer relative to the other along the armchair direction (Fig. 1(a)). It also explores the Coulomb excitations observed in the electron energy loss spectrum (EELS), resulting from the vertical/non-vertical Dirac-cone structures, highly distorted parabolic dispersions, unusual Fermi-momentum and band-edge states, and particular distribution of subenvelope functions on four sublattices. Moreover, we evaluates whether single-particle excitation channels exist between or within pairs of energy bands at different doping levels and bilayer configurations.
The plasmon dispersions of acoustic and optical modes are analyzed with respect to different Landau dampings and critical transferred momenta, and the (${q}$, $\omega$)-phase diagrams at different Fermi levels are evaluated to display the single-particle and undamped collective excitations in the entire region.
These results have the potential to provide insight into the impact of interlayer coupling on high-resolution EELS\cite{NanoLetters14;3827,Carbon114;70} and inelastic light scattering spectroscopy\cite{PRL98;206802,PRL111;106801} performed on 2D materials.

\section{Tight-binding model and modified RPA}

When an electron beam is incident on bilayer graphene with uniform charge density, the time-dependent external Coulomb potential causes charge density fluctuations on each graphene layer. Electron-electron scattering in different directions of the incident electron beam satisfies energy conservation and momentum conservation. Additionally, the charge redistribution results in induced and effective Coulomb potentials on the graphene layers. These potentials can be described and incorporated into Dyson's equation using linear response theory and the layer-based RPA\cite{PRB34;979,PLA352;446}. The exact expression for the effective Coulomb potential between the $l$- and $l^\prime$-th layers, under screening effects due to $\pi$ and $\pi^\star$ electrons, is given by:

\begin{equation}
\epsilon_{0}V^{eff}_{ll^{\prime}}(\mathbf{q},\omega)=V_{ll^{\prime}}(\mathbf{q})+
\sum\limits_{mm^{\prime}}V_{lm}(\mathbf{q})P_{mm^{\prime}}(\mathbf{q},\omega)
V^{eff}_{m^{\prime}l^{\prime}}(\mathbf{q},\omega)\text{,}
\end{equation}%
where $P_{mm^{\prime}}$ is the bare polarization function ascribed as the linear coefficient of the response theory,
\begin{equation}
\begin{array}{l}
P_{mm^{\prime}}(\mathbf{q},\omega)=2\sum\limits_{k}\sum\limits_{h,h^{\prime}=c,v}\sum\limits_{n,n^{\prime}}
\biggl(\sum\limits_{s}
U_{smh}(\mathbf{k})
U^{\star}_{sm^{\prime}h^{\prime}}(\mathbf{k+q})\biggr)\\
\times\biggl(\sum\limits_{s}
U^{\star}_{smh}(\mathbf{k})
U_{sm^{\prime}h^{\prime}}
(\mathbf{k+q})\biggr)\times\frac{f(E^{h}_{n}(\mathbf{k}))-f(E^{h^{\prime}}_{n^{\prime}}(\mathbf{k+q}))}
{E^{h}_{n}(\mathbf{k})-E^{h^{\prime}}_{n^{\prime}}(\mathbf{k+q})
+\hbar\omega+i\Gamma}
\text{.}
\end{array}
\end{equation}%

The equation (1) includes two terms, the first of which is the bare Coulomb potential, denoted as $V_{ll}^{\prime}=v_{q}e^{-q|l-l^{\prime}|d_{0}}$, where $d_{0}$ is the distance between two graphene layers. This term is characterized by the Coulomb potential of the 2D electron gas, $v_q=2\pi e^2/q$, where $q$ is the transferred momentum during scattering. The second term, the induced potential, is proportional to the total screening charge density, which is obtained by solving the Poisson equation. This term is also proportional to the effective potential in the self-consistent approach. To account for the effects of band-structure, the layer-decomposed wave functions scheme is utilized. The function $P$ in equation (2) involves all the layer-dependent electron-hole excitations between the initial and final states, denoted as $E^{h}_{n}(\mathbf{k})$ and $E^{h^{\prime}}_{n^{\prime}}(\mathbf{k+q})$ respectively. The subscript $c$ ($v$) indicates valence (conduction) bands with the band index $n$ ($n^{\prime}$). The prefactor 2 accounts for spin degeneracy, and $\Gamma$ represents the energy width due to deexcitation mechanisms. The amplitude product $UU^{\ast}$ is used to calculate the expectation value in the long wavelength limit. Moreover, the Fermi-Dirac distribution function $f(E^{h}_{n}(\mathbf{k}))=1/[1+\exp{(E^{h}_{n}(\mathbf{k})-\mu(T)})/k_{B}T]$ is utilized, where $k_{B}T$ and $\mu(T)$ represent the Boltzmann constant and chemical potential respectively. By using equation (1) and (2), the effective potential $V^{eff}$ resulting from the Coulomb perturbation on the available excitation channels appearing on different graphene layers can be obtained.

The dimensionless energy loss function, $\mathbf{Im}[-1/\epsilon]$, being directly proportional to the measured EELS intensity, is defined as follows under the Born approximation\cite{PLA352;446},

\begin{equation}
\begin{split}
\mathbf{Im}[-1/\epsilon]&\equiv\sum\limits_{l}\mathbf{Im}\biggl[-V_{ll}(\mathbf{q},\omega) \biggr]
/\biggl(\sum\limits_{lm}V_{lm}(\mathbf{q})/N\biggl)\text{.}
\end{split}
\end{equation}%
The denominator in this equation represents the average of the external potentials across all graphene layers, making Eq. (3) an appropriate description for emergent layered systems, such as group-IV\cite{NJP16;125002} and group-V 2D\cite{PRB98;115411} materials. By including charge screening, this equation provides comprehensive information on various plasmon modes, while Eq. (2) describes only single-particle (electron-hole) excitations. To study single-particle electronic properties, we employ the tight-binding model to calculate the eigen-energies and eigen-functions.  Specifically, one of the graphene layers is continuously shifted with ${\delta}$ along the armchair direction ($\widehat{x}$), creating AA (${\delta}$=0), AB (${\delta}$=6/8), and AA$^{\prime}$ (${\delta}$=12/8) configurations (Fig. 1).

The Hamiltonian for the $\pi$-electrons, which is constructed using the four 2p$_z$ orbitals in a primitive unit cell, is given by
\begin{equation}
H=-\sum_{i,j}\gamma_{ij}c_{i}^{\dag}c_{j}\text{.}
\end{equation}
Here, $\gamma_{ij}$ denotes the hopping integral between the $i$-th and $j$-th lattice sites connected by vector ${\bf d}$, and is well-fitted to describe the 2$p_z$-orbital interactions as a function of ${\bf d}$.\cite{SciRep4;7509} This approach has been successfully applied to various sp$^2$ systems related to graphene that exhibit significant layer-layer interactions, such as multi-walled carbon nanotubes \cite{JPSJ68;3806} and multilayer graphenes\cite{PCCP17;26008}. The hopping parameter $\gamma_{ij}$ has an analytic form and is given by

\begin{equation}
\gamma_{ij}=\gamma_{0}e^{-\frac{d-b_{0}}{\rho}}\left[1-\left(\frac{\mathbf{d}\cdot\mathbf{e_{z}}}{d}\right)^{2}\right]
+\gamma_{1}e^{-\frac{d-d_{0}}{\rho}}\left(\frac{\mathbf{d}\cdot\mathbf{e_{z}}}{d}\right)^{2}\text{,}
\end{equation}
where ${\gamma_0=-2.7}$ eV and ${\gamma_1=0.48}$ eV represent the hoppings between nearest intralayer and vertical interlayer sites, respectively. Interlayer distance $d_{0}=3.35$ $\AA$ and decay length $\rho=0.184b$ are used to cut off the interaction. Although VASP calculations indicate that AB stacking is the most stable configuration, the slight variation in interlayer distance across different configurations is negligible\cite{JPCC119;10623}. It should be noted that the electronic model for the anisotropic effects has been generalized to investigate the magnetic quantization, with a particular focus on the diverse Landau levels and their corresponding optical selection rules.\cite{SciRep4;7509}

\section{Energy band structures and DOS}

The stacking symmetry in bilayer graphene directly affects its electronic properties. In Figs. 1(b)-1(g), the band structures were analyzed for various ${\delta}$ values, with a focus on the low-lying energy bands shown in the zoomed-in 3D plot. The Fermi level is indicated by black dashed lines at zero energy ($E_{F}$=0). The inversion symmetry maintains the equivalence of the two low-lying valleys, K and K$^\prime$, for all systems. The interlayer hopping integral causes a notable change in the carrier concentration, resulting in semiconductors, semimetals, and metals. The AA bilayer with $\delta$=0 displays the first and second pairs of Dirac cones ($1^{c,v}$ and $2^{c,v}$) at the K/K$^\prime$ point, each with Dirac-point energies of ${E^{c,v}}$=0.40 eV and -0.45 eV, respectively, as shown by the black curves in Fig. 1(b). The significant overlap of the two vertical Dirac cones indicates the presence of high-concentration free electrons and holes. The AB bilayer with $\delta=1$ shows two pairs of parabolic bands, with a valence-conduction ($1^{c}-1^{v}$) band intersection point at the Fermi level, belonging to the zero-gap semimetal, as shown by the black curves in Fig. 1(c). However, the relative shift in bilayer graphene breaks the symmetry of the band structure. As ${\delta}$ gradually increases from 0 to 1/8, the energy dispersion deviates from the linear dispersion, as indicated by the red curves in Fig. 1(b). The 3D band structures reveal that the strong hybridization of the two cones along ${\hat k_x}$ results in a closed region without states around $E_{F}=0$ (Fig. 1(d)). It is situated between the edges of the conduction arc and the valence arc, forming two saddle points at the top and bottom of the arc-shaped region without states.

As the bilayer graphene undergoes a further shift towards the AB-stacked configuration, there is a significant change in its band structure. The arc-shaped region expands gradually, with highly distorted energy dispersions observed at $\delta$=4/8 and $\delta$=6/8 (Figs. 1(e) and 1(f)), and the enclosed stateless volume disappears completely at $\delta$=1 due to the interlayer atomic interactions. However, the conservation of electronic states leads to the transfer of the massless Dirac states near the K point to separated saddle points in its neighboring region that belong to the Fermi momentum states. This dominance of low-lying electronic properties persists during the transition from AA to AB stacking. Furthermore, the complete separation of two pairs of energy bands indicates the reformation of two individual systems with band-edge states located at ${E^{c,v}\sim0}$ and ${E^c\sim0.4}$ eV and ${E^v\sim-0.45}$ eV, respectively. These characteristics are expected to be crucial for electronic excitation. The low-lying band structures of sliding bilayer graphene also undergo dramatic transformations as the relative shift changes from $\delta$=1 to 12/8. The parabolic bands of the AB stacking are strongly distorted along ${\hat k_y}$ and ${-\hat k_y}$ simultaneously, resulting in the creation of two new Dirac points outside the arc region that are significantly tilted up and down at the K point due to the strong hybridization with energies of the order of $\gamma_{1}$ for the two neighboring conduction (valence) bands. Most importantly, as shown in the case of AA$^\prime$ stacking with $\delta=12/8$ in Fig. 1(g), two isotropic Dirac cones are formed and lie in opposite directions for the conduction and valence bands. These characteristics can be verified by tracing the trajectories of constant-energy loops measured from the two separated Dirac cones.

Figure 2 clearly shows the diverse electronic structures as vHSs in the DOS. Two types of fascinating structures are observed across $E_{F}$=0 in the low-energy DOS, namely plateau and dip structures, due to electronic hybridization. The former is attributed to the linear energy dispersion in the AA and AA$^{\prime}$ stacking configurations (Figs. 2(a) and 2(f)). On the other hand, the dip structure arises from the arc-shaped stateless region (Figs. 2(b)-2(d)) in low-symmetry systems deviating from the typical AA, AB, and AA$^{\prime}$ stackings, where the energy of the dip corresponds to that of the conduction-valence band intersection\cite{IOPBook;978}. Parabolic bands corresponding to AB-stacked or near-AB-stacked configurations contribute to the formation of asymmetric peaks in the DOS (blue circles in Figs. 2(c)-2(e)). Moreover, the low-symmetry systems exhibit symmetric peaks at energies about $\pm0.2$ eV (red circles), induced by the new saddle points of the arc-shaped pockets depicted in Fig. 1. The number and energy of these peaks are sensitive to the relative shift. The DOS intensity is proportional to the free carrier densities and can be used to evaluate the intensity of electronic excitations in the sliding bilayer graphene system.


\section{Diverse Coulomb excitations in pristine systems and doping-enriched excitation phenomena}

Sliding bilayer systems possess unique electronic properties that give rise to various Coulomb excitation phenomena. The single-particle polarization functions described in Eq. (2) are composed of four layer-specific components: ${P_{11}}$, ${P_{22}}$, ${P_{12}}$, and $P_{21}$. The inversion symmetry of the bilayer system guarantees that ${P_{11}}$=${P_{22}}$ and ${P_{12}}$=${P_{21}}$. The spectral structures of ${P_{11}}$ and ${P_{12}}$ are determined by the four atoms $A_{1}$, $A_{2}$, $B_{1}$, and $B_{2}$ in the primitive unit cell. We provide a detailed analysis of the frequency response of the polarization for the variations of $\delta$, $E_{F}$, and ${q}$, illustrated in Figs. 3 and 4, which plot the real and imaginary parts of the independent polarization functions. The excitation channels enable us to identify the intensity and form of the polarization functions based on the energy dispersions, dimensionalities, and Kramers-Kronig relations\cite{PLA352;446}. Moreover, the classification of electron-hole excitations and the Landau dampings of the plasmon modes are facilitated by vHSs, Fermi-momentum states, and the critical points that arise from band-edge states and saddle points in the energy-wavevector space.

The bare polarization functions of undoped AA-stacked bilayer graphene exhibit distinct structures due to the presence of free carriers. This is clearly illustrated in Figs. 3(a)-3(d) by the black curves. The imaginary part of ${P_{11}}$ and ${P_{12}}$ exhibit a first square-root asymmetric peak at ${\omega_{-}\sim3\gamma_0bq/2}$ in the form of ${1/\sqrt {\omega_{-}^{2}-\omega^{2}}}$ (indicated by the inverted triangle symbol), while the Kramers-Kronig relation shows the real part to be divergent in the opposite square-root form. This peak arises from intrapair/intraband transitions, specifically ${1^{v}\rightarrow1^{v}}$ and ${2^{c}\rightarrow2^{c}}$, as indicated by the arrow in the insert. These transitions occur at Fermi-momentum states with linear dispersions. Such excitations can survive at transferred momenta comparable to the Fermi momentum. However, interpair transitions are prohibited within a wide frequency range of ${\omega_{-}<\omega<\omega_{+}}$ (${\sim0.1-0.85}$ eV), as demonstrated by the vanishing intensity in the single-particle electronic polarizations, i.e., ${\mathbf{Im}[P_{11}]}$ and ${\mathbf{Im}[P_{12}]}$=0. This absence of single-particle excitations facilitates the formation of undamped 2D intraband plasmon modes in the Dirac cone. The other prominent excitations are identified as ${1^{v}\rightarrow1^{c}}$ and ${2^{v}\rightarrow2^{c}}$, which contribute to four ${1/\sqrt {\omega^{2}-\omega_{+}^{2}}}$ asymmetric peaks for ${\omega>0.85}$ eV, where $\omega_{+}$ is the interband excitation frequency. Moreover, the main characteristics of the response function also depend on the magnitude of ${q}$, such as ${q=0.02}$ ${\AA^{-1}}$ (green curve). The dependence of the wavefunction and DOS on $\mathbf{k}$ suggests that intraband/interband excitations cause a relatively stronger/weaker response. It is worth noting that the asymmetric Dirac cone allows for the possibility of two peaks for a specific interband excitation channel with a small $\omega$ difference relative to the K/K$^\prime$ point. Unlike previous work that only demonstrated one peak\cite{AOP339;298}, the formation of multiple peaks is a result of the model in which the interlayer hopping integral considers not only the vertical sites but also the non-vertical sites.

For a deviation of $\delta=1/8$ from AA-stacking in bilayer graphene, the single-particle polarization functions still display one- and four-peak structures associated with intraband and interband excitations within the intrapair, as shown by the red curves in Fig. 3(a). This indicates that the system with $\delta=1/8$ differs from the AA-stacked system due to the extra vHSs near $E_F=0$ that arise from the deviation in stacking. However, the system with $\delta=1/8$ still exhibits a similar excitation behavior to the AA-stacked system due to the linear dispersion of energy that is present in both systems. There is a significant decrease in the intensity of the first asymmetric peak at $\omega_{-}$ due to the reduction of total free carriers in the arc-shaped region, which weakens the Coulomb response. The interband electronic excitations of the second structure are also distributed over a wider $\omega$ range but with less intensity, mainly due to the enhanced asymmetry of the energy spectrum and reduced DOS. It is important to note that the vHSs of the saddle points induce a prominent intensity of optical absorption but do not create any special structures in the bare polarization functions\cite{SciRep4;7509}. The differences between the two different types of excitations reflect their respective many-particle and single-particle excitation mechanisms of the Coulomb field and the electromagnetic field.

Compared to the AA stack configuration, the AB under layer shift results in a dramatic transformation in the electronic properties. The polarization function of the normally stacked AB graphene is compact, with only two types of interband excitations, as shown by the black curves in Figs. 3(e)-3(h). The symmetry of the wavefunction leads to forbidden transitions in the interpair channel. At ${\omega\sim0.005}$ eV, the interband excitations from parabolic dispersion, ${1^{v}\rightarrow1^{c}}$, induce a shoulder structure of the imaginary part and a logarithmic structure of the real part. Similarly, the polarization function contributed by ${2^{v}\rightarrow2^{c}}$ induces similar behavior at ${\omega\sim0.91}$ eV. Specifically, in semimetallic AB bilayer graphene, it may be difficult to observe prominent peaks in the electron energy loss spectrum. This is because plasmon modes are expected to be suppressed, and only electron-hole excitations exist in the ($q$, $\omega$)-phase diagram. However, the sliding system, in general, can be viewed as an integration of abundant single-particle excitations, exhibiting extra plasmon modes and rather strong Landau dampings throughout the diagram.

In bilayer graphene with a ${\delta=6/8}$, shown by the red curves in Figs. 3(e)-3(h), the combination of the four subenvelope functions of the first and second graphene layers results in more available excitation channels. These non-vertical Coulomb excitations, which include interpair transitions ${1^{v}\rightarrow2^{c}}$ and ${2^{v}\rightarrow1^{c}}$, in addition to intrapair transitions, produce three distinct features in the polarization function. These three structures conform to the Kramers-Kronig relation for a complex response function, which is calculated by integrating the principle-value of $\omega$ for linear Fermi-momentum states, saddle-point states, and band-edge states\cite{PRB34;979}. The strong threshold response at around ${0.022}$ eV for a specific ${q=0.005}$ ${\AA^{-1}}$ is attributed to electron-electron interaction of the Fermi momentum states. Moreover, saddle-point transition generates the second symmetry/shoulder structure at around ${0.24}$ eV, which exhibits a logarithmic/discontinuous form in the imaginary/real parts of the polarization function. As the excitation energy increases, interpair excitations appear in the range of 0.45 eV ${<\omega<}$ 0.62 eV, revealing the third special structure. Finally, the band-edge states of the second parabolic valence and conduction bands ($2^{c,v}$) can produce prominent shoulders and logarithmic peaks at around 0.88 eV in the imaginary- and real-part bare response functions, respectively. For larger values of ${q}$, such as ${q=0.02}$${\AA^{-1}}$, the components of the mixed excitation channels can be identified. However, the abundant single-particle excitation activity would pose a significant obstacle to the generation of undamped plasma modes.

Doping effects bring about significant diversity in both single- and many-particle electronic excitations. When the rigid band is shifted up or down, the doping effect causes the excited electrons and excited holes to become more asymmetric about the Fermi level from the viewpoint of Coulomb excitation. The increase in free carriers greatly diversifies the excitation channels, resulting in a dramatic expansion, shift, and transformation of the peak structure of the polarization function with changes in the Fermi surface. We can describe the three categories of excitation frequency ranges as follows. To illustrate the doping effects on the response functions, we consider the cases of ${\delta=6/8}$ and ${\delta=1}$ as model materials. Figures 4(a)-4(d) show the two excitation channels caused by the non-crossing bands, namely intrapair intraband and interband transitions (${1^{c}\rightarrow1^{c}}$, ${1^{v}\rightarrow1^{c}}$ and ${1^{c}\rightarrow2^{c}}$), which dominate the low-energy excitation, rather than the original excitation associated with the Dirac point at $E_{F}=0$. The first square-root peak is enhanced at ${\omega\sim3\gamma_0bq/2}$, directly reflecting the robustness of the Dirac excitations remaining at different $\delta$$^{'}$s. In the range of 0.2 eV ${<\omega<}$ 0.6 eV, extra structures appear with a strong response, while the highest channel at ${\omega\sim0.88}$ eV, classified as the third category, remains largely unchanged until the doping concentrations reach the $2^{c}$ energy band. However, in terms of the Dirac-cone excitations for AA stacking and its counterpart, the unmixed wavefunction under the rigid-band approximation is responsible for the interpair excitation that disappears in this range under doping. As for the mixed states (Fig. 1(g)), the distribution of excitations associated with linear dispersion on $\omega$ would become wider as $q$ and $E_{F}$ increase.

In the last stage of $\delta=12/8$, the two Dirac cones that are not vertically aligned at the K point result in well-behaved wavefunctions with linear superpositions of four subenvelope functions. As seen in Figs. 4(e)-4(h), the imaginary-part response function displays additional square-root asymmetric peaks at frequencies of around 0.029 eV, 0.37 eV, 0.68 eV, and 0.82 eV. These peaks arise from the Fermi-momentum states of ${1^{v}\rightarrow1^{v}}$ (${2^{c}\rightarrow2^{c}}$), ${2^{v}\rightarrow1^{c}}$, ${1^{v}\rightarrow1^{c}}$, and ${2^{v}\rightarrow2^{c}}$. The interpair interband channel of ${1^{c}\rightarrow2^{c}}$ is forbidden due to the well-behaved functions in the linear superposition of the four subenvelope functions. In addition, there are no excitation splittings in the higher-frequency channels, since the parabolic conduction/valence band disappears in the first/second Dirac cone. Further deviation of the geometric symmetry, such as $\delta=11/8$, also results in unique Coulomb excitations in Figs. 4(a)-4(d), which are as complex as those in the former two cases.

Regarding the $\omega$-range for single-particle excitations, special structures at frequencies of around 0.029 eV, 0.30 eV to 0.32 eV, and 0.66 eV to 0.83 eV are contributed by all intrapair transitions (${1^{v}\rightarrow1^{v}}$ and ${2^{c}\rightarrow2^{c}}$), interpair ones (${2^{v}\rightarrow1^{c}}$ and ${2^{c}\rightarrow1^{c}}$), and intrapair ones (${1^{v}\rightarrow1^{c}}$ and ${2^{v}\rightarrow2^{c}}$), respectively. The distortion of the non-vertical Dirac cones induced by $\delta$, which is accompanied by higher-energy parabolic dispersions, is responsible for the specific divergent structure. It should be noted that the threshold peak remains for $q$ in the order of Fermi momentum, while the second structure becomes rather asymmetric due to the additional ${2^{c}\rightarrow1^{c}}$ channel. The extra channels of ${1^{v}\rightarrow2^{c}}$ and ${1^{c}\rightarrow2^{c}}$ cannot generate any special structures in the undoped case because of the absence of Fermi-momentum states and parabolic band-edge states for electronic transitions.

Moreover, for the third region, the higher-frequency polarization function is dominated by intrapair interband transitions, showing two neighboring prominent peaks of the specific channels of ${1^{v}\rightarrow1^{c}}$ and ${2^{v}\rightarrow2^{c}}$. The splitting excitation phenomena purely come from highly hybridized energy bands since the conduction/valence band of the first/second pair has separated Dirac-cone and parabolic dispersions. In summary, all sliding bilayer graphenes illustrate a close relationship between excitation channels and $E_F$ in terms of polarization. The intensity, number, form, and frequency of the special structures are highly sensitive to changes in free carrier densities. Furthermore, the conversion of single-particle excitation channels under the influence of the Coulomb field helps understand the mechanism of collective excitation of screened electrons.

The energy loss spectral function defined in Eq. 4 reveals various screened phenomena where the 2D plasmon modes can be damped by different electron-hole pair excitations. In the case of the pristine AA stacking, Fig. 5(a) demonstrates two significant peaks at ${\omega\sim}$0.22 eV and 1.09 eV under ${q=0.005}$ ${\AA^{-1}}$, with the former being particularly prominent. The first plasmon mode arises due to intrapair intraband excitations, with the excitation frequency significantly higher than the single-particle frequency $\sim$0.028 eV in Fig. 3. This plasmon represents the collective charge oscillations of free electrons and holes solely induced by the strong interlayer hopping integrals in the high-symmetric AA stacking configuration. Collective excitations do not experience any Landau damping, given the absence of both intrapair interband and interpair interband transitions at small transferred momenta. Another higher-frequency plasmon mode arises from intrapair interband channels, but their strength is weakened by the single-particle excitations (inset in Fig. 5(a)). However, the plasmon peak intensities are evidently reduced in the ${\delta=1/8}$ stacking (Fig. 5(b)). This result directly reflects the low-lying distorted Dirac-cone structures with the lower free carrier density and the stronger Landau damping. Regarding the plasmon frequencies, only a slight enhancement/decrease is observed in the first/second mode. With a further increase in shift, such plasmon modes demonstrate dramatic changes in their intensities. They are relatively weak in the $\delta=6/8$ and $\delta=1$ stackings (Figs. 5(c) and 5(d)) and may be challenging to observe in the energy loss spectra. The main reason comes from the low DOS intensity for Fermi-momentum states/free carriers. Finally, the second plasmon peak almost disappears for the ${\delta=11/8}$ and ${\delta=12/8}$ systems (Figs. 5(e) and 5(f)). The intensity of the first plasmon mode is significantly lower than that in the AA case, but it is notably higher than in the AB case. The non-vertical Dirac-cone band structure shown in Fig. 2(g) is responsible for these results, as it creates additional excitation channels and plasmon excitations.

Electron or hole doping can result in unique and intricate phenomena in the plasmon peaks of energy loss spectra, owing to the heightened presence of free carriers in the asymmetric energy spectra. In the AA stacking depicted in Fig. 5(a), the frequency and strength of the two plasmon modes do not exhibit a simple and monotonic relationship with an increase in the Fermi surface. Particularly, the first plasmon mode remains unchanged at ${E_F}$=0.2 eV, but shows weakened intensity at 0.4 eV. This is because the Fermi level is situated above and near the Dirac point with nearby linear dispersion, causing non-uniform distributions of free electrons and holes. Similar results are observed in the ${\delta=1/8}$ stacking, as illustrated in Fig. 5(b). However, in the ${\delta=6/8}$ stacking, the first plasmon mode is significantly strengthened, and is somewhat prominent in the AB stacking, accompanied by the emergence of the second peak. In these two systems, the increase in free conduction electrons is responsible for the enhanced plasmon excitations. Lastly, doping in the ${\delta=11/8}$ and ${\delta=12/8}$ stackings leads to an increase in the intensity and frequency of the first plasmon mode, and this effect is much stronger than that observed in the AA stacking. However, increasing $E_{F}$ cannot generate the second plasmon mode due to the independence of the two non-vertical Dirac cones. As the transferred momentum ${q}$ increases, the plasmon oscillation typically experiences a blue shift, and the Landau damping effect gradually becomes evident. The actual dispersion relation can be illustrated in the (${q}$ , $\omega$)-phase diagram, which is constructed by incorporating both single-particle and many-particle excitations.


The (${q}$, $\omega$)-phase diagrams of sliding bilayer systems reveal the rich plasmon features resulting from various Coulomb excitation phenomena, which are influenced by geometry and doping, as shown in Figs. 6-8. In Fig. 6,
it is evident that the electronic excitations of undoped graphene are sensitive to stacking configurations and low-lying energy bands. Layer shifts can significantly affect bright plasmon modes and electron-hole excitation boundaries. The dashed and solid lines indicate the lower bounds for energy and momentum conservation. These bounds may change with $\delta$ due to modifications in the electronic structure. As shown in Figs. 6(a) and 6(b), the bright intensity in the diagrams indicates that the two types of collective excitations can only survive simultaneously in the $\delta=0$ and $\delta=1/8$ systems. The first and second plasmons are classified as the acoustic and optical modes, respectively\cite{AOP339;298}. However, they undergo different magnitudes of Landau damping from single-particle excitations. The former has a $\sqrt{q}$-relation at small momenta, similar to that of a 2D electron gas\cite{PRL18;546}. Both free electrons and holes in two distinct Dirac cones contribute to the collective charge oscillations at long wavelengths. This plasmon has no damping within $q<0.05$ $\AA^{-1}$, but experiences strong intrapair interband damping after a critical momentum of $q_c>0.1$ $\AA^{-1}$, making the energy loss relatively difficult to observe. In particular, for the $\delta=6/8$ and $\delta=1$ stackings, the two plasmon modes almost disappear, as shown in Figs. 6(c) and 6(d), except for the faint acoustic mode at very small $q$ due to the low free carrier density. In contrast, only the first plasmon mode with lower intensity is observed at $q_c\sim0.05$ $\AA^{-1}$ in the $\delta=11/8$ and $\delta=12/8$ bilayer graphenes in Figs. 6(e) and 6(f). That is to say, the AA$^\prime$ stacking exhibits significant differences from the AA system in the distribution regions for electron-hole and plasmon excitations, mainly attributed to the tilted alignment of the two Dirac cones and the energy spacing for carrier excitations.

The increment of the free carriers leads to significant changes of the many-particle effects on the Coulomb excitation spectra, as shown in Figs. 7 and 8, where the electron-hole boundaries also vary with $E_F$, as indicated by the dashed and solid curves. For the AA and ${\delta=1/8}$ bilayer stackings, an increase in $E_F$ causes a non-homogeneous distribution of the same carrier density of free electrons and holes, plotted in the band structures in Figs. 2(d). As a result, the efficiency of collective charge oscillations is hindered, reducing the plasmon strength in the screened response spectrum. The undamped momentum range and critical transferred momentum of the acoustic plasmon mode decrease, leading to significant changes of loss spectra under different $E_F$ values, e.g., ${E_F}$=0, 0.2 eV and 0.4 eV, shown in Figs. 6(a)-6(b), Figs. 7(a)-7(b), and 8(a)-8(b). However, adjusting $E_F$ significantly enhances the carrier density of free conduction electrons in the ${\delta=6/8}$ and AB systems. The first plasmon mode exists at any doping levels (Figs. 7(c)-7(d) and Fig. 8(c) and 8(d)), while the second one appears only at Fermi levels sufficiently high enough to trigger major interband excitations, e.g., ${E_F}$=0.4 eV in Figs. 8(c) and 8(d). There is an absence of a simple relation between the acoustic plasmon strength and ${E_F}$ for these two types of stackings. The $E_F$-enhanced e-e interaction effects on the energy bands of the ${\delta=11/8}$ and ${\delta=12/8}$ bilayer stackings greatly strengthen the collective oscillation for the acoustic plasmon, while they cannot create a higher-frequency optical plasmon. It is worth mentioning that the acoustic plasmon mode shows a significant frequency increase as the Fermi level increases. This mode is observable at moderate Fermi levels, as it is sufficiently strong. However, it may be weakened by Landau damping from specific single-particle excitations.


\section{Concluding Remarks}

In summary, we utilize the generalized tight-binding model to investigate the energy loss spectra of sliding bilayer graphene systems with various stacking shifts. The findings shed light on the plasmon modes and collective charge oscillations of free electrons and holes that are induced by interlayer hopping and Coulomb interactions. Compared with the electromagnetic wave perturbation, the full information of the bare response function is caused by the Coulomb perturbation, mainly due to the non-perpendicular excitation of the displacement-enriched energy band. The (${q}$, $\omega$)-phase diagrams reveal the screened phenomena that lead to damped 2D plasmon modes in response to changes in the stacking shift and Fermi level, with the doping concentration having a significant impact, as discussed. Two significant plasmon modes were observed in the pristine AA stacking configuration. However, the plasmon peak intensities are reduced with the variation of $\delta$, reflecting the distorted Dirac-cone structures with the lower free carrier density and stronger Landau damping. Doping can result in unique and intricate phenomena in the plasmon peaks of energy loss spectra. Furthermore, the non-uniform distributions of excitations near the Dirac point with distorted energy dispersion cause the plasmon modes to be weakened or strengthened with respect to different $\delta$. Overall, this study provides a fundamental understanding of the plasmon features of sliding bilayer graphene systems and paves the way for potential implications in the design of graphene-based plasmonic devices\cite{NatNanotechnol6;630,SCIENCE344;1369}.

\section{Acknowledgments}
This work was supported in part by the National Science Council of Taiwan, the Republic of China, under Grant Nos. NSC 105-2112-M-006 -002 -MY3 and National Research Foundation of Korea (201500000002559).

\newpage
\renewcommand{\baselinestretch}{0.2}

\newpage
\centerline {\textbf {Figure Captions}}

Figure 1: (a) Geometric structures of sliding bilayer graphene with different relative shifts $\delta$ along the armchair direction, where $d_{0}$ indicates the layer distance. Figures 1(b)-(g) show the corresponding band structures and close-ups of the 3D band structure near the Fermi level for different $\delta$ values.

Figure 2: Low-energy density of states for different $\delta$ values in sliding bilayer graphene: (a) ${\delta=0}$, (b) ${\delta=1/8}$, (c) ${\delta=6/8}$, (d) ${\delta=1}$, (e) ${\delta=11/8}$, and (f) ${\delta=12/8}$ (AA$^\prime$). The circles indicate the vHSs of the DOS.

Figure 3: Real and imaginary parts of the two independent bare polarizations at two different $q$ values, $q=0.005$ ${1/\AA}$ and of $q=0.02$ ${1/\AA}$, and at $E_F=0$ for the ${\delta=0}$ (AA) and ${\delta=1/8}$ bilayer stacking: (a) $\mathbf{Re}[P_{11}]$, (b) $\mathbf{Im}[P_{11}]$, (c) $\mathbf{Re}[P_{12}]$, and $\mathbf{Im}[P_{11}]$, where the real part and imaginary part are represented by the bold $\mathbf{Re}$ and $\mathbf{Im}$. Similar plots are shown in (e)-(h) for the $\delta=1$ and $\delta=6/8$ systems. The insets in the figure provide a close-up view of the first divergent structure, with the corresponding excitation channel illustrated in (b).

Figure 4: Similar bare response functions at different Fermi levels for the (a)-(d) $\delta=6/8$ and $\delta=1$ systems, and for the (e)-(h) $\delta=11/8$ and $\delta=12/8$ systems under same variables as in Fig. 3.

Figure 5: Electron energy loss spectra for different $E_F$ values and $q$ values for sliding bilayer systems with different $\delta$ values. Also shown in the insets are close-ups near the higher-frequency plasmon.

Figure 6: ($q$, $\omega$)-phase diagrams of the pristine sliding bilayer graphenes for different $\delta$: (a) ${\delta=0}$, (b) ${\delta=1/8}$, (c) ${\delta=6/8}$, (d) ${\delta=1}$, (e) ${\delta=11/8}$, and (f) ${\delta=12/8}$.

Figure 7: Similar plots to Fig. 6 are shown, indicating the distinct doping effects for ${E_F=0.2}$ eV.

Figure 8: Similar plots to Fig. 6 are shown, indicating the distinct doping effects for ${E_F=0.4}$ eV.

\begin{figure}
\centering
\includegraphics[width=0.9\linewidth]{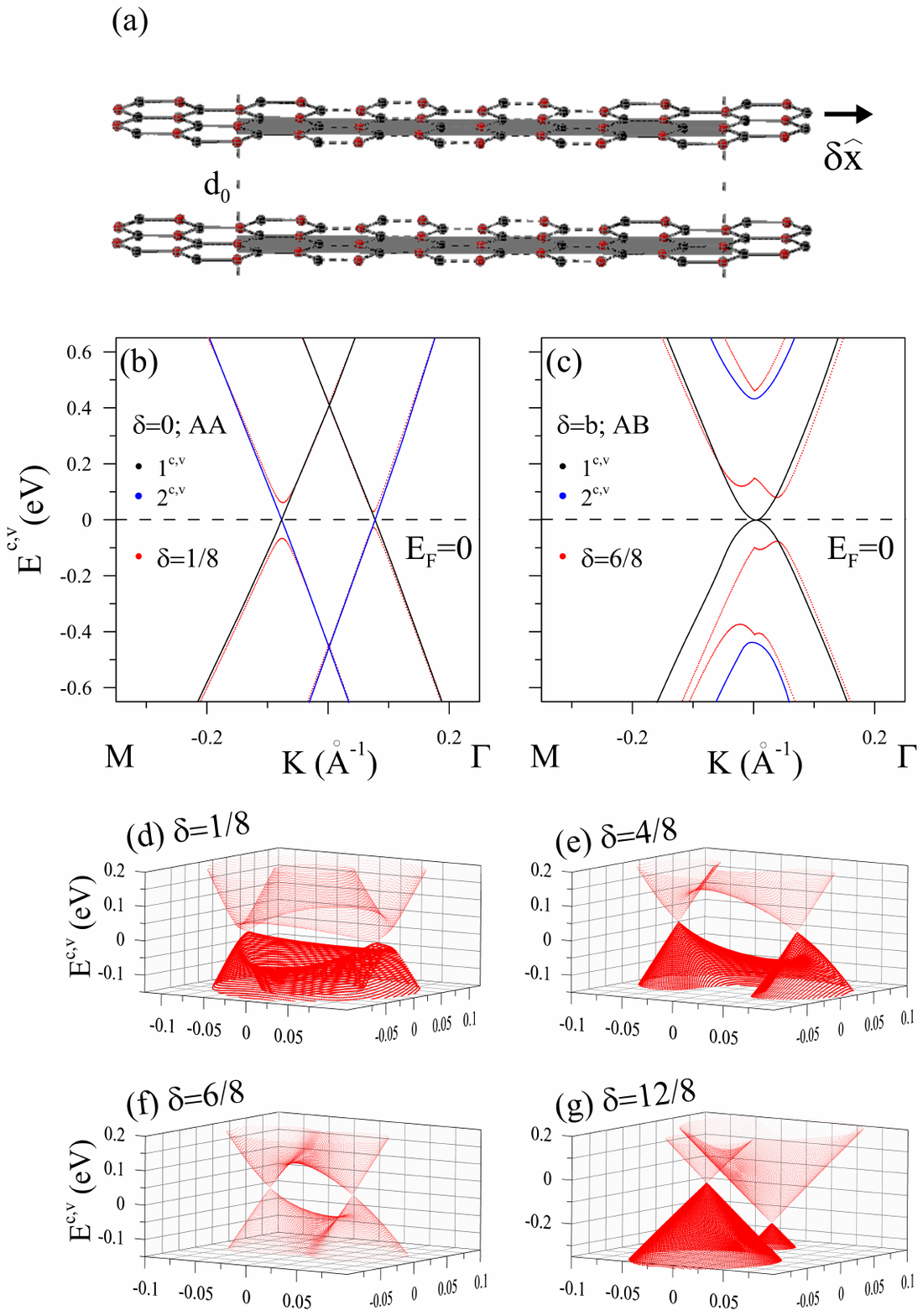}
\caption{(a) Geometric structures of sliding bilayer graphene with different relative shifts $\delta$ along the armchair direction, where $d_{0}$ indicates the layer distance. Figures 1(b)-(g) show the corresponding band structures and close-ups of the 3D band structure near the Fermi level for different $\delta$ values.}
\label{fig:graph}
\end{figure}

\begin{figure}
\centering
\includegraphics[width=0.9\linewidth]{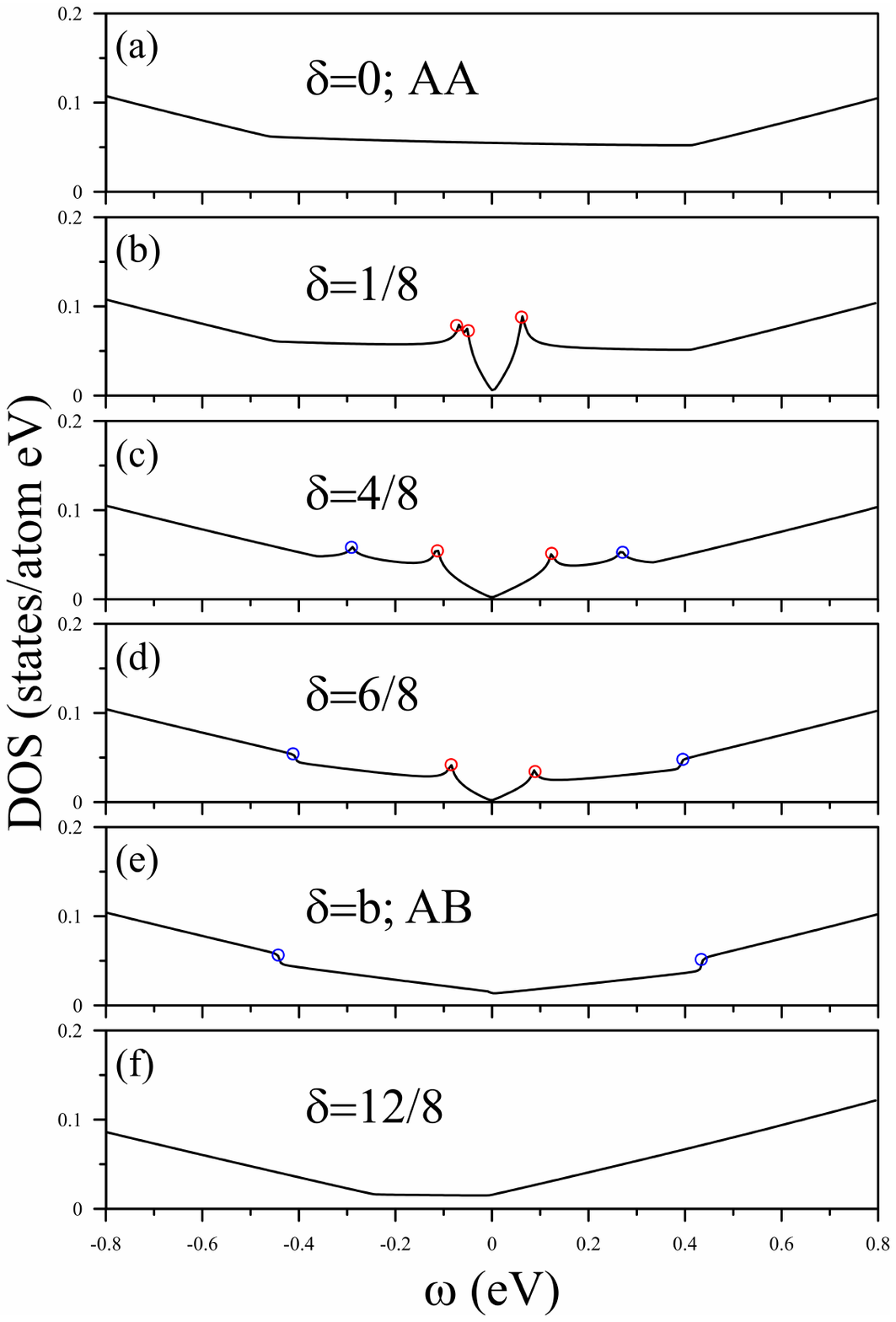}
\caption{Low-energy density of states for different $\delta$ values in sliding bilayer graphene: (a) ${\delta=0}$, (b) ${\delta=1/8}$, (c) ${\delta=6/8}$, (d) ${\delta=1}$, (e) ${\delta=11/8}$, and (f) ${\delta=12/8}$ (AA$^\prime$). The circles indicate the vHSs of the DOS.}
\label{fig:graph}
\end{figure}

\begin{figure}
\centering
\includegraphics[width=0.9\linewidth]{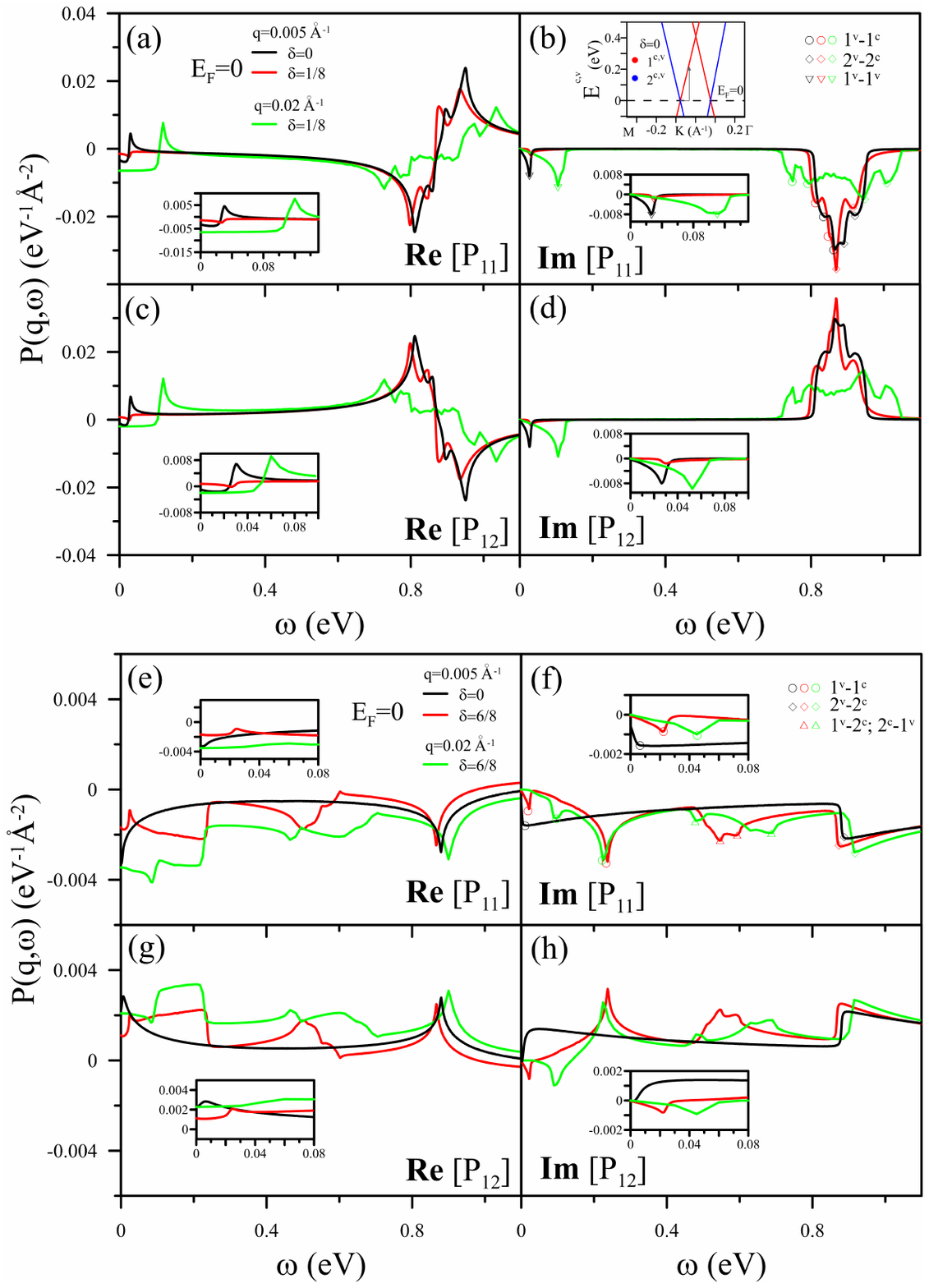}
\caption{Real and imaginary parts of the two independent bare polarizations at two different $q$ values, $q=0.005$ ${1/\AA}$ and of $q=0.02$ ${1/\AA}$, and at $E_F=0$ for the ${\delta=0}$ (AA) and ${\delta=1/8}$ bilayer stacking: (a) $\mathbf{Re}[P_{11}]$, (b) $\mathbf{Im}[P_{11}]$, (c) $\mathbf{Re}[P_{12}]$, and $\mathbf{Im}[P_{11}]$, where the real part and imaginary part are represented by the bold $\mathbf{Re}$ and $\mathbf{Im}$. Similar plots are shown in (e)-(h) for the $\delta=1$ and $\delta=6/8$ systems. The insets in the figure provide a close-up view of the first divergent structure, with the corresponding excitation channel illustrated in (b).}
\label{fig:graph}
\end{figure}

\begin{figure}
\centering
\includegraphics[width=0.9\linewidth]{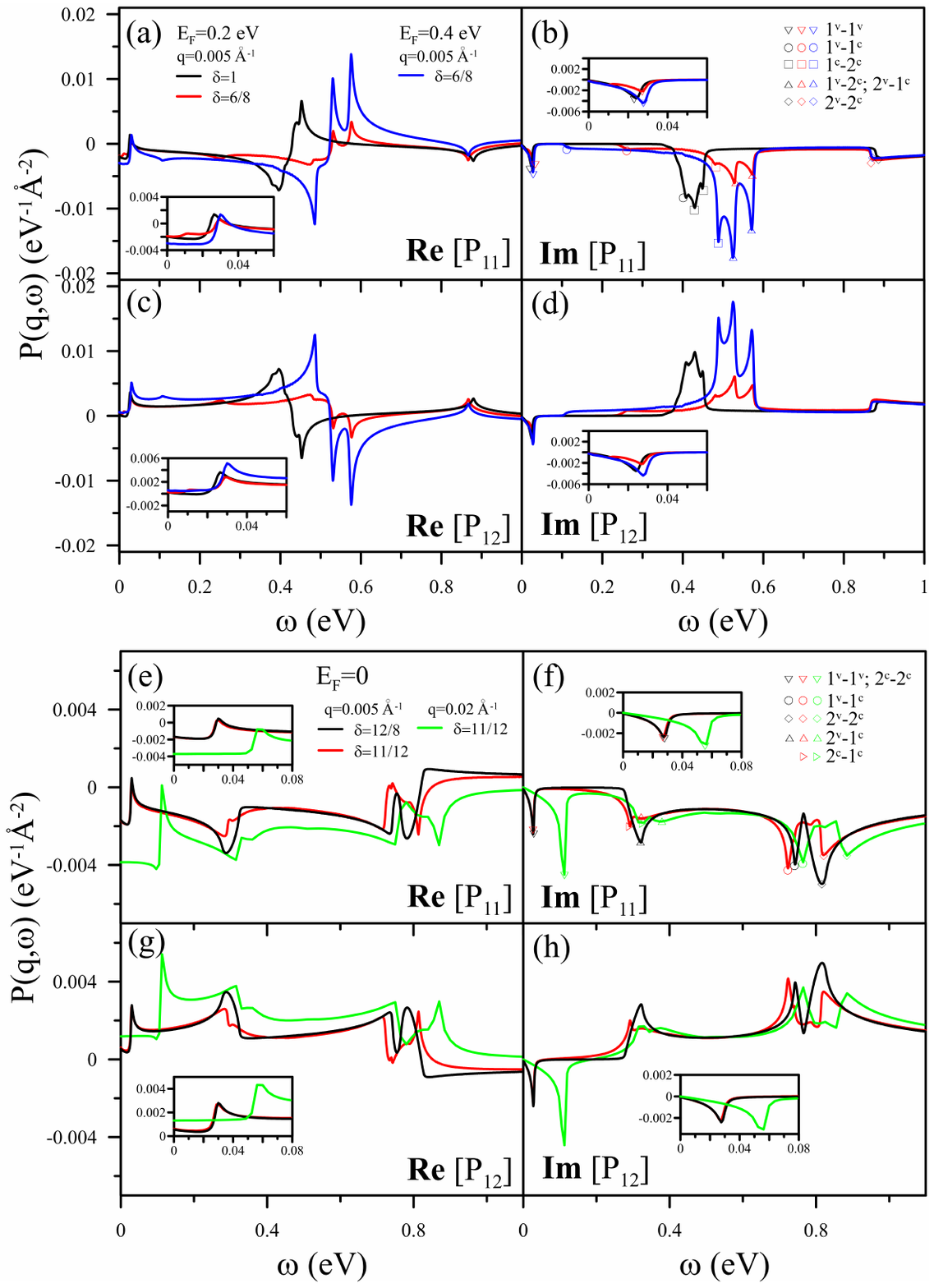}
\caption{Similar bare response functions at different Fermi levels for the (a)-(d) $\delta=6/8$ and $\delta=1$ systems, and for the (e)-(h) $\delta=11/8$ and $\delta=12/8$ systems under same variables as in Fig. 3.}
\label{fig:graph}
\end{figure}

\begin{figure}
\centering
\includegraphics[width=0.9\linewidth]{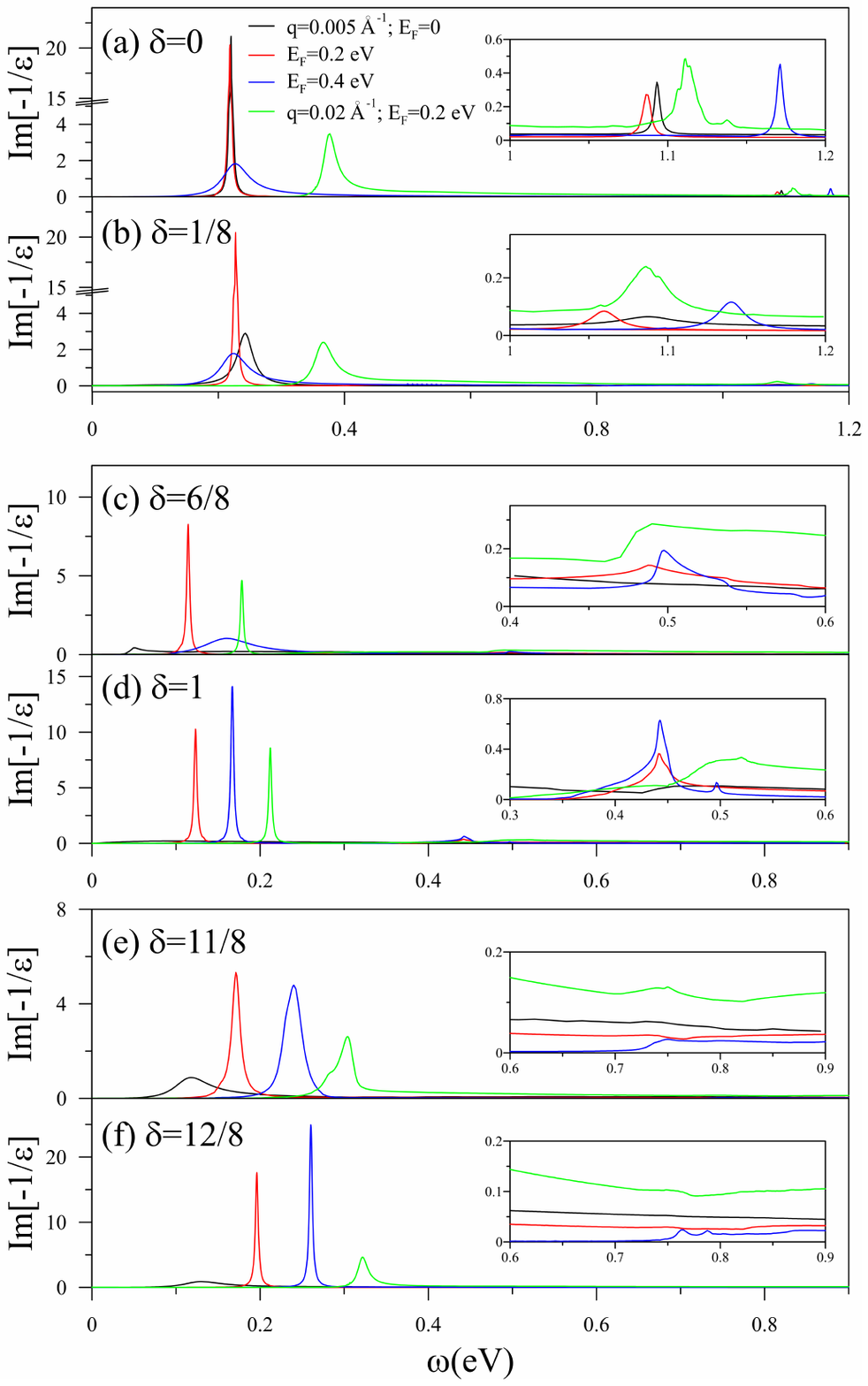}
\caption{Electron energy loss spectra for different $E_F$ values and $q$ values for sliding bilayer systems with different $\delta$ values. Also shown in the insets are close-ups near the higher-frequency plasmon.}
\label{fig:graph}
\end{figure}

\begin{figure}
\centering
\includegraphics[width=0.9\linewidth]{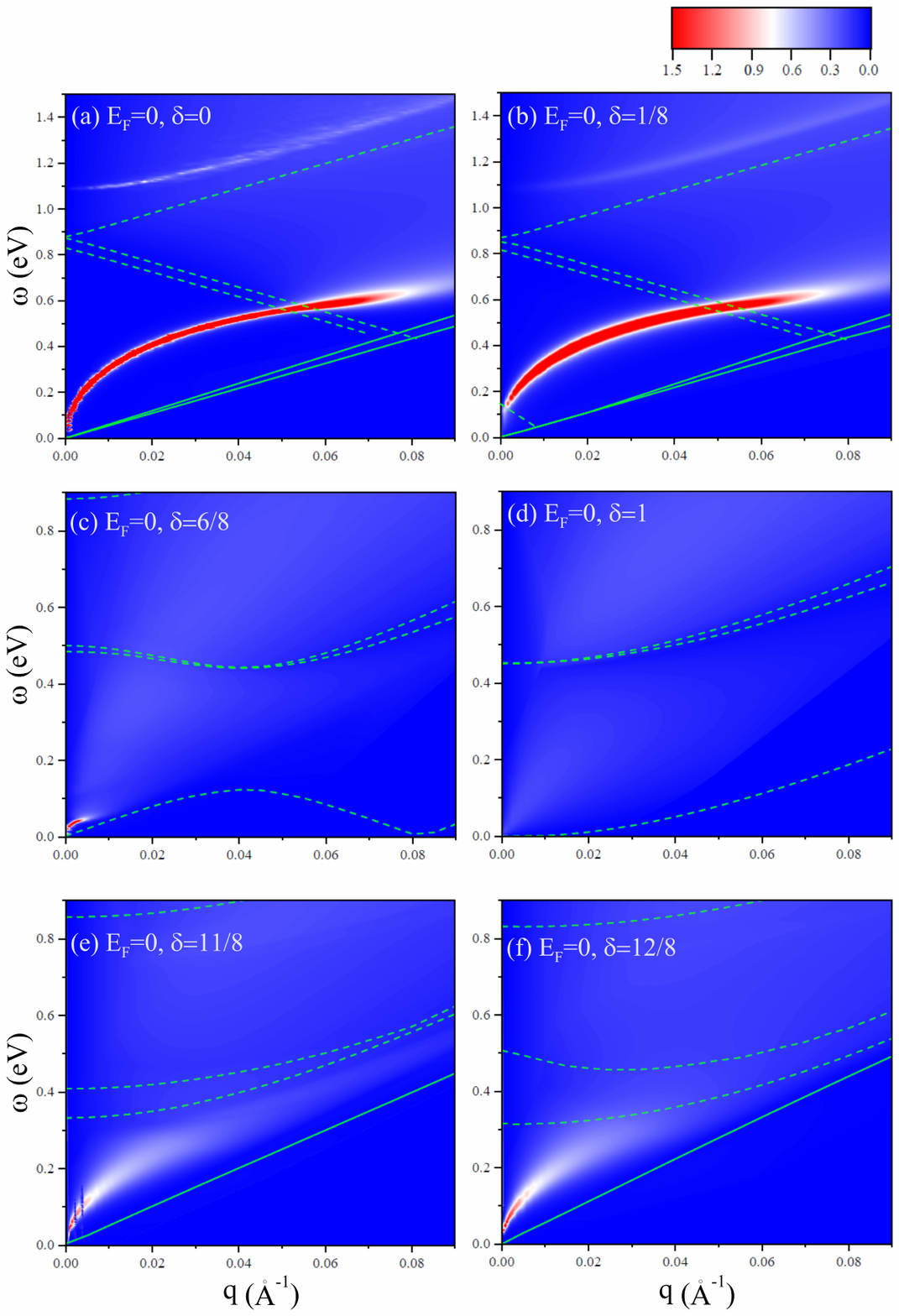}
\caption{($q$, $\omega$)-phase diagrams of the pristine sliding bilayer graphenes for different $\delta$: (a) ${\delta=0}$, (b) ${\delta=1/8}$, (c) ${\delta=6/8}$, (d) ${\delta=1}$, (e) ${\delta=11/8}$, and (f) ${\delta=12/8}$.}
\label{fig:graph}
\end{figure}

\begin{figure}
\centering
\includegraphics[width=0.9\linewidth]{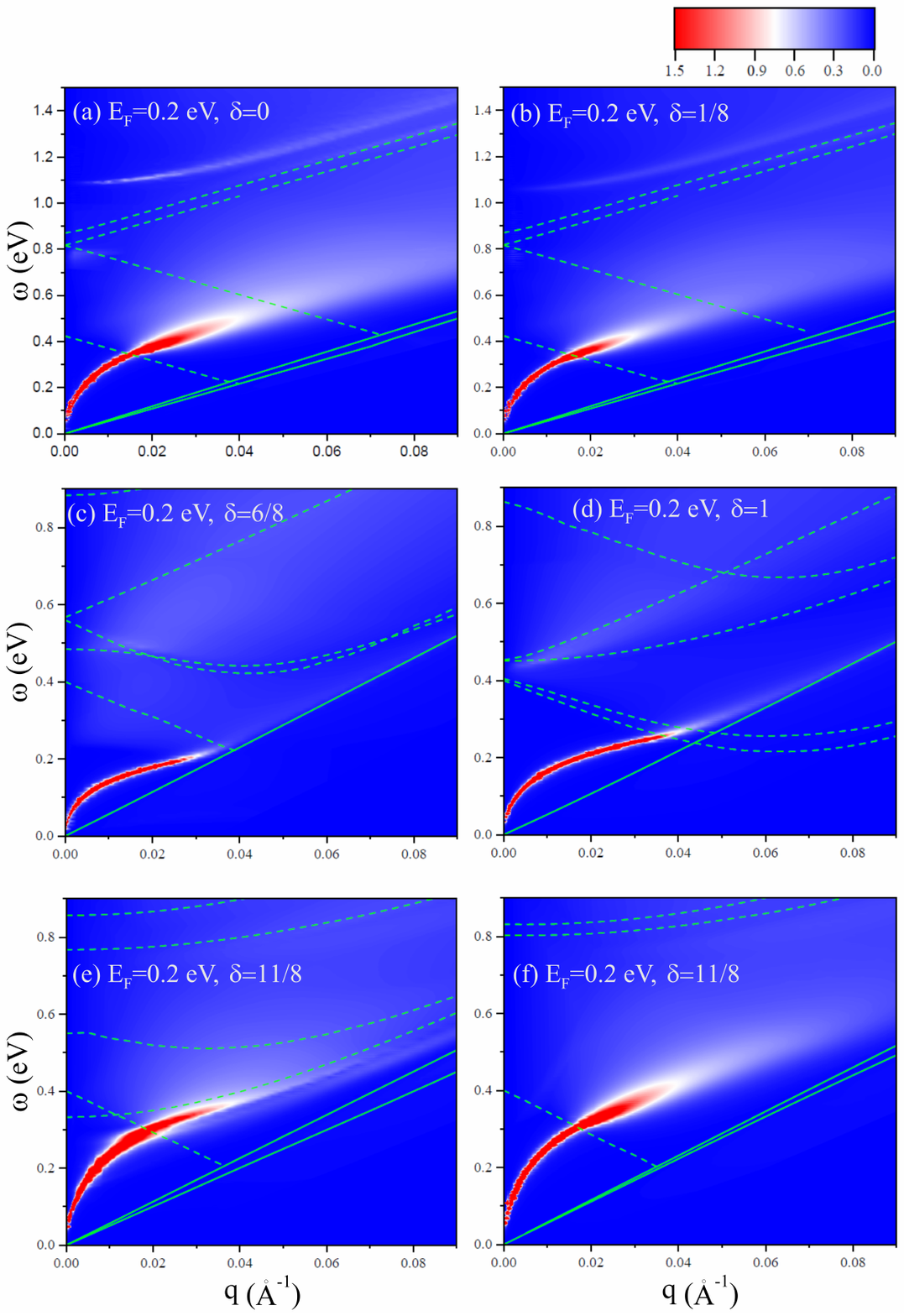}
\caption{Similar plots to Fig. 6 are shown, indicating the distinct doping effects for ${E_F=0.2}$ eV.
}
\label{fig:graph}
\end{figure}

\begin{figure}
\centering
\includegraphics[width=0.9\linewidth]{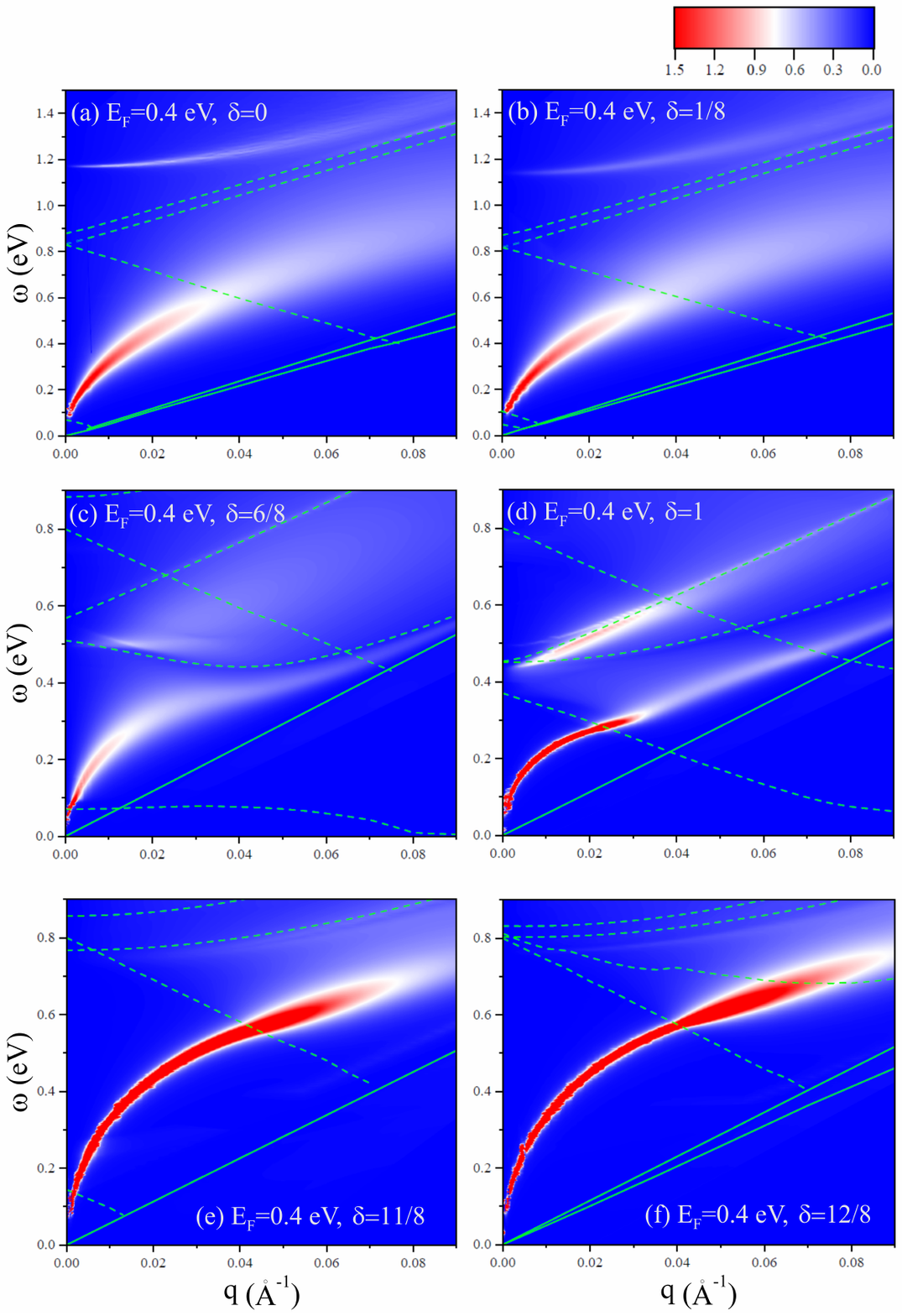}
\caption{Similar plots to Fig. 6 are shown, indicating the distinct doping effects for ${E_F=0.4}$ eV.}
\label{fig:graph}
\end{figure}

\end{document}